\documentclass[12pt]{article}
\usepackage[mathscr]{eucal}
\usepackage{amsmath,amsfonts,amssymb,amsthm}
\usepackage[matrix,arrow]{xy}
\usepackage{soul,xcolor}
\setstcolor{red}

\newdimen\tableauside\tableauside=1.0ex
\newdimen\tableaurule\tableaurule=0.4pt
\newdimen\tableaustep
\def\phantomhrule#1{\hbox{\vbox to0pt{\hrule height\tableaurule
width#1\vss}}}
\def\phantomvrule#1{\vbox{\hbox to0pt{\vrule width\tableaurule
height#1\hss}}}
\def\sqr{\vbox{%
  \phantomhrule\tableaustep

\hbox{\phantomvrule\tableaustep\kern\tableaustep\phantomvrule\tableaustep}%
  \hbox{\vbox{\phantomhrule\tableauside}\kern-\tableaurule}}}
\def\squares#1{\hbox{\count0=#1\noindent\loop\sqr
  \advance\count0 by-1 \ifnum\count0>0\repeat}}
\def\tableau#1{\vcenter{\offinterlineskip
  \tableaustep=\tableauside\advance\tableaustep by-\tableaurule
  \kern\normallineskip\hbox
    {\kern\normallineskip\vbox
      {\gettableau#1 0 }%
     \kern\normallineskip\kern\tableaurule}%
  \kern\normallineskip\kern\tableaurule}}
\def\gettableau#1 {\ifnum#1=0\let\next=\null\else
  \squares{#1}\let\next=\gettableau\fi\next}

\newtheorem{prop}{Proposition}[section]

\theoremstyle{definition}

\newcommand{\bref}[1]{\textbf{\ref{#1}}}

\def\be{\begin{equation}}
\def\ee{\end{equation}}
\def\ba{\begin{array}}
\def\ea{\end{array}}

\def\dps{\displaystyle}

\def\fnote#1#2{\begingroup\def\thefootnote{#1}\footnote{#2}\addtocounter
{footnote}{-1}\endgroup}
\def\cH{\mathcal{H}}

\def\cM{\mathcal{M}}

\def\cW{\mathcal{W}}

\numberwithin{equation}{section} \makeatletter
\@addtoreset{equation}{section}

\begin{document}

\begin{flushright}
FIAN-TD-2014-08 \\
\end{flushright}

\bigskip
\bigskip
\begin{center}

{\Large\textbf{Conformal blocks of $\cW_N$ Minimal Models\\ \vspace{4mm}
and AGT correspondence}}

\vspace{.5cm}

{\large{K.B. Alkalaev} $^{1}$\fnote{$\dagger$}{email: alkalaev@lpi.ru} ,
{\large V.A. Belavin} $^{1,2}${\,}\fnote{*}{email: belavin@lpi.ru}}

\bigskip
\bigskip

\begin{tabular}{ll}
$^{1}$~\parbox[t]{0.9\textwidth}{\normalsize\raggedright
{\it I.E. Tamm Department of Theoretical Physics, P.N. Lebedev Physical Institute, Leninsky ave. 53,
119991 Moscow, Russia}}\\
$^{2}$~\parbox[t]{0.9\textwidth}{\normalsize\raggedright
{\it Department of Quantum Physics, Institute for Information Transmission Problems, Bolshoy Karetny per. 19, 127994 Moscow, Russia}}
\end{tabular}\\

\vspace{5mm}

\begin{abstract}
We study  the AGT correspondence between four-dimensional supersymmetric gauge field theory and two-dimensional conformal field theories
in the context of $\cW_N$ minimal models. The origin of the AGT correspondence  is in a
special integrable structure which appears in the properly extended conformal theory.
One of the basic manifestations of this integrability is the special orthogonal basis
which arises in the extended theory. We propose modification of the AGT
representation for the $\cW_N$ conformal blocks in the minimal models.
The necessary modification is related to the reduction of the orthogonal basis.
This leads to the explicit combinatorial representation for the conformal blocks
of minimal models and employs the sum over N-tupels of Young diagrams with
additional restrictions.

\end{abstract}

%\vspace{5mm}
\end{center}

\newpage
{
\footnotesize
%\scriptsize
\tableofcontents
}

\section{Introduction}

%\subsection{Some preliminaries}

The bootstrap approach to 2d CFTs   is based on the
requirements of  conformal symmetry, associativity of the
operator algebra and crossing symmetry for the correlation
functions \cite{BPZ}. One of the main ingredients in the conformal bootstrap
approach is the conformal block function, which sums up the
contributions of  conformal descendants of a given primary
field. The bootstrap  approach allows in principle to define the
structure constants of the operator algebra and then to construct
arbitrary multi-point correlation functions. Important class of
the conformal field theories is the minimal models, where there is
a finite number of  irreducible representations of the
conformal algebra closed with respect to the operator algebra.

The AGT correspondence \cite{AGT} and its generalizations
establish connections between different 2d conformal field
theories and instanton moduli spaces in 4d $\mathcal{N} = 2$
supersymmetric gauge quiver theories in the Omega background
\cite{Nekrasov}. In the framework of this correspondence the
conformal blocks are represented  by the instanton partition
functions which are known  explicitly \cite{Nekrasov}. The AGT
representation  for $\cW_N$ conformal blocks was considered in
Refs. \cite{Wyl,Mironov:2009qt,Mironov:2009by}. In Ref. \cite{Litvinov}  the
connection between an integrable structure of the theory with chiral algebra
\begin{equation}
\mathcal{A}_N=
\mathcal{H}
\otimes \frac{\widehat{\mathfrak{sl}}(N)_1 \otimes \widehat{\mathfrak{sl}}(N)_{n-1}}{\widehat{\mathfrak{sl}}(N)_n}\;,
\end{equation}
and $\cW_N$ conformal blocks was found.
Here $\mathcal{H}$ denotes the
Heisenberg algebra and the second term gives the standard coset realisation of the $\cW_N$ algebra with
the central charge defined in terms of the parameter $n$. The theory possesses the following remarkable
property. The algebra $\mathcal{A}_N$ is some special limit of the
quantum toroidal $\mathbf{gl}_{1}(\mathrm{q}_1,\mathrm{q}_2)$ algebra with
simple action on the cohomologies of equivariant K theories. In
particular, in the module of the toroidal algebra  there exists
the orthogonal basis enumerated by plane partitions.
$\mathcal{A}_N$ module inherits  the structure of the basis of the
module of the toroidal algebra \cite{Belavin:2011pp}. As a result
of the limiting procedure the basis is  enumerated  by $N$ ordinary
Young diagrams. In the $\mathcal{A}_N$ module there exists some
special orthogonal basis  such that the matrix element of the
composite vertex operators in this basis are known explicitly in
terms of the simple rational functions of the basic parameters
\cite{AlbaLitvinov,Litvinov}. For this reason, using the
orthogonal basis automatically leads to the explicit results for
arbitrary correlation functions.

It is interesting to study the consequences of the AGT
correspondence for the conformal blocks of $\cW_N$ minimal models.
The first example of the AGT correspondence for rational CFT models with Virasoro symmetry 
was considered  in Ref. \cite{Santachiara:2010bt}.
In  \cite{Estienne:2011qk} 
the properties of the conformal blocks of some special degenerate fields in the conformal field theories
with extended $\cW_N$ symmetry were studied in the context of the AGT correspondence. However, the problem
of constructing the general combinatorial representations
for conformal blocks in $\cW_N$ minimal models and more generally  the problem of constructing the correlation functions 
remains open.  
Interesting applications of $\cW_N$ minimal  models, where the question of constructing conformal blocks is relevant, can be found
within the $AdS_3/CFT_2$ higher spin correspondence, see,
\textit{e.g.}, Refs. \cite{Papadodimas:2011pf,Chang:2011vk}.

In this paper we propose AGT-like combinatorial representation for
the conformal block functions  in $\cW_N$ minimal models. We
formulate a necessary modification of the AGT combinatorial
representation for the Virasoro minimal models. The modification
is reduced to additional restrictions on the region of
summation over Young diagrams. In particular, we perform some checks for Virasoro
conformal blocks comparing with the exact results which follows
from the definition of the conformal block. Also, we formulate the
conjecture on the form of AGT representation for conformal
blocks of $\cW_N$ minimal models. Our conjecture relies upon the properties
of the orthogonal basis in $\mathcal{A}_N$
modules described  in Refs. \cite{FFJMM,FFJMM2}.

\section{AGT for non-degenerate Virasoro  representations }

As an example of the AGT representation for non-degenerate representations of  Virasoro  algebra $\cW_2$
we consider a $4$--point conformal block on the sphere. The consideration of  general $k$--point correlation functions
contains the same ingredients. The general answer including degenerate representations of $\cW_N$ algebras  will be
given in Section  \bref{sec:W_N}.

The 4--point conformal block is a
holomorphic contribution of the conformal family $[\Phi_{\Delta_0}]$ of the primary field $\Phi_{\Delta_0}$ in the correlation function
$\langle\Phi_{\Delta_1}(x)\Phi_{\Delta_2}(0)\Phi_{\Delta_3}(1)\Phi_{\Delta_4}(\infty)\rangle$.
A standard Liouville parametrization of the conformal dimensions and central charge
is
\begin{equation}
\label{LiouvParam}
\Delta_i=Q^2/4-P_i^2,\qquad c=1+6 Q^2, \qquad Q=b^{-1}+b\,.
\end{equation}
The AGT correspondence gives the following power series
expansion for the 4--point conformal block \cite{AGT}
\begin{equation}
\label{4pointBlock}
\mathcal{B}(P_i;x)\equiv \sum_{N=0}^{\infty} x^N \mathcal{B}^{(N)}(P_i)=(1-x)^{-\nu}\sum_{N=0}^{\infty}\!
x^{N} F^{(N)}\,,
\end{equation}
and
\begin{equation}
\label{Dc}
F^{(N)}=\sum_{\vec \lambda,|\vec \lambda|=N}\frac{Z_{f}(\mu_i,a;\vec \lambda)}{Z_{v}(a;\vec \lambda)}\,.
\end{equation}
The summation on the right hand side runs over pairs of Young diagrams $\vec \lambda=(\lambda_1,\lambda_2)$
and the norm $|\vec \lambda|$ denotes  the total number of cells. The explicit form of
$Z_{f}(\mu_i,a;\vec \lambda)$ and $Z_{v}(a;\vec \lambda)$ reads
\begin{equation}
\label{Zf}
\begin{aligned}
Z_{f}(\mu_i,a;\vec \lambda)=
\prod_{s\in \lambda_{1}}(\phi(a,s)+\mu_1)(\phi(a,s)+\mu_2)(\phi(a,s)+\mu_3)(\phi(a,s)+\mu_4)\\
\times\prod_{s\in \lambda_{2}}(\phi(-a,s)+\mu_1)(\phi(-a,s)+\mu_2)(\phi(-a,s)+\mu_3)(\phi(-a,s)+\mu_4)\,,
\end{aligned}\end{equation}
and
\begin{equation}
\label{Zv}
\begin{aligned}[centered]
Z_{v}(a;\vec \lambda)=
\prod_{s\in \lambda_{1}}E_{\lambda_1,\lambda_2}\!\bigl(2a\bigl|s\bigr)\big(Q-E_{\lambda_1,\lambda_2}\!\bigl(2a\bigl|s\bigr)\big)
E_{\lambda_1,\lambda_1}\!\bigl(0\bigl|s\bigr)\big(Q-E_{\lambda_1,\lambda_1}\!\bigl(0\bigl|s\bigr)\big)\\
\times\prod_{s\in \lambda_{2}}E_{\lambda_2,\lambda_1}\!\bigl(-2a\bigl|s\bigr)\big(Q-E_{\lambda_2,\lambda_1}\!\bigl(-2a\bigl|s\bigr)\big)
E_{\lambda_2,\lambda_2}\!\bigl(0\bigl|s\bigr)\big(Q-E_{\lambda_2,\lambda_2}\!\bigl(0\bigl|s\bigr)\big)\,.
\end{aligned}\end{equation}
Functions $E_{\lambda,\mu}\!(x\bigl|s\bigr)$ and $\phi(x,s)$ are defined as
\begin{equation}
\label{E-def}
\ba{l}
E_{\lambda,\mu}\bigl(x\bigl|s\bigr)=x-b\,l_{\scriptscriptstyle{\mu}}(s)+b^{-1}(a_{\scriptscriptstyle{\lambda}}(s)+1)\;,
\\
\\
\phi(x,s) = x+ b (i-1)+b^{-1}(j-1)\;.

\ea
\end{equation}
To explain our notation
we adjust $(\lambda)_i$ to $i$th row of the Young diagram $\lambda$
and  denote $(\lambda)^T_j$ the length of the $j$th column, where $T$ stands for a matrix transposition.
For a cell $s=(i,j)$ such that $i$ and $j$ label a respective row  and a column,
the arm-length function $a_{\scriptscriptstyle{\lambda}}(s)$ and the leg-length function
$l_{\scriptscriptstyle{\lambda}}(s)$ are given by
\be
\label{legsarms}
a_{\scriptscriptstyle{\lambda}}(s) = (\lambda)_i-j\;,
\qquad
l_{\scriptscriptstyle{\lambda}}(s) = (\lambda)^T_j - i\;.
\ee
The parameters of Nekrasov partition function
are related to the parameters of the conformal block as follows
\begin{eqnarray}\label{AGTparam}
&\mu_{1}=\frac Q2-(P_1+P_2),\quad \mu_2=\frac Q2-(P_1-P_2),\, \nonumber\\
&\mu_3=\frac Q2-(P_3+P_4),\quad \mu_4=\frac Q2-(P_3-P_4),\, \\
&a=P_0,\quad \nu=2(\frac{Q}{2}-P_1)(\frac{Q}{2}-P_3)\nonumber\,.
\end{eqnarray}

Following Ref. \cite{AlbaLitvinov} one can find an orthogonal basis
$|P,\vec{\lambda}\rangle$ numerated by pairs of Young diagrams in
$\mathcal{A}_N$ module that reproduces
Nekrasov decomposition \eqref{4pointBlock}-\eqref{Dc}. The
Nekrasov sum is obtained simply by inserting the following unity decomposition
\be
\sum_{\vec{\lambda}} \frac{|P, \vec{\lambda}\rangle \langle P, \vec{\lambda}|}
{\langle P, \vec{\lambda}| P, \vec{\lambda}\rangle} = \mathbb{I}\;
\ee
in the correlation function  $\langle\Phi_{\Delta_1}(x)\Phi_{\Delta_2}(0)\Phi_{\Delta_3}(1)\Phi_{\Delta_4}(\infty)\rangle$ between each two of the primary fields.

\section{The AGT-like representation  for the $\cW_2$  minimal models}

Conformal field theories ${{\cal M}}_{p,p'}$
are characterized by the central charge of the Virasoro algebra
\begin{equation}
\label{CenChaCFT}
c=1-6\frac{(p'-p)^2}{pp'}\,,
\qquad
b=i\sqrt{\frac{p'}{p}}\,.
\end{equation}
There are $(p-1)\times (p'-1)/2$ primary fields $\Phi_{l,k}$
($l=1,\ldots, p-1$ and $k=1,\ldots,p'-1$)
in the model. The conformal dimensions  are determined by the Kac formula
\begin{equation}
\label{SpectCFT}
{
\Delta_{m,n}=\frac{(p'm-p n)^2-(p'-p)^2}{4pp'}\,.
}
\end{equation}
Note that there is a symmetry $\Delta_{m,n} = \Delta_{p-m, p^\prime - n} $ and  $\Delta_{m,n} = \Delta_{p+m, p^\prime +n} $.
In the Liouville parametrization the values of the parameter $P$
corresponding to the degenerate values are
\begin{equation}
\label{Liouv2}
P_{m,n}= P(\Delta_{m,n})=\frac{m b+n b^{-1}}{2}\,.
\end{equation}
For the Virasoro minimal models the fusion rules
which describe conformal families that appear in the operator
product expansion of two primary fields are \cite{BPZ}
\begin{equation}
\label{VirFusion}
\Phi_{(r,s)}\otimes \Phi_{(m,n)}=
\sum_{\substack{k=|m-r|+1,\\ k-m+r-1 \,\,\text{even}}}^{\text{min}(m+r-1,2p'-1-m-r)}\;
\sum_{\substack{l=|n-s|+1,\\l-n+s-1\,\,\text{even}}}^{\text{min}(n+s-1,2p-1-n-s)}\;
\left[ \Phi_{(k,l)} \right]\,.
\end{equation}

\subsection{Reduction of the basis in the minimal models}
\label{sec:3.1}

The AGT representation \eqref{4pointBlock} is not directly applicable to minimal models.
The reason is that the fields of minimal models are degenerate.
Indeed, vectors of invariant submodules possess
zero norms so that  expression \eqref{4pointBlock} which contains
these norms in the denominator is singular in this case.
One comment about relations between general and minimal models conformal block
is in order. The 4--point conformal blocks on the sphere in $\cM_{p,p^\prime}$ minimal model can be derived
from the expression for the non-degenerate fields by means of the following procedure
of the analytic continuation.

Let us consider first CFT for general value of the central charge parameter, and
let
\be
\langle\Phi_{\Delta_1}(x) \Phi_{\Delta_2}(0) \Phi_{\Delta_3}(1) \Phi_{\Delta_4}(\infty)\rangle
\ee
be the correlation function of four non-degenerate fields. Suppose we have
some orthogonal basis in the Verma module $[\Phi_{\Delta}]$ denoted by
$|N\rangle$.
For the conformal block we have
\be\label{Block}
\mathcal{B}(x)=\sum_{N=0}^{\infty}\, x^N \,\frac{\langle \Phi_{\Delta_1} \Phi_{\Delta_2}|N\rangle \langle N|\Phi_{\Delta_3} \Phi_{\Delta_4}\rangle}{\langle N|N\rangle}\;.
\ee
Now, we are interested in the conformal blocks of minimal models. They can be derived
from \eqref{Block} in two steps. First, we fix $c=c_{p,p'}$ \eqref{CenChaCFT} and  external
dimensions $\Delta_{i}=\Delta_{m_i,n_i}(p,p')$ \eqref{SpectCFT}. We note that the set of the dimensions
should be admissible for the fusion rules of the minimal models.
Second, we take a limit
$\Delta\rightarrow\Delta_{mn}$. More precisely, we use the parametrization $\Delta=
a(Q-a)$ and take $a\rightarrow a_{mn}$.
 Among the descendants of the primary
field $\Phi_{\Delta}$ will appear singular vectors (and their descendants).
Let us denote the singular vector creation operator as $D_{mn}$.
One can derive \cite{AlZ} that the norm of the vector $D_{mn} \Phi_{\Delta(a)}$
has zero of the first order in the limit $a\rightarrow a_{mn}$, or, explicitly,
\be
\langle D_{mn}  \Phi_{\Delta(a)} |D_{mn}  \Phi_{\Delta(a)}\rangle\sim (a-a_{mn})\;.
\ee
One can check that each of the three-point functions in the numerator of \eqref{Block} has also zero of
the first order in this limit as it was also  shown in Ref. \cite{AlZ}.
Hence we get a second-order zero in the numerator and a first-order zero in the denominator.  The same result occurs for all
descendants of the basic
singular vectors $D_{mn} \Phi_{\Delta(a)}$ and $D_{p-m,p'-n} \Phi_{\Delta(a)}$. So we can just drop out the contribution of the vectors
in the decomposition \eqref{Block} which fall in the invariant
subspace generated by the singular vector $D_{mn} \Phi_{\Delta(a)}$.
In fact, this procedure can be considered as a definition of the conformal blocks
of the minimal models.

On the other hand, the above procedure can be effectively used  to re-derive
conformal blocks in minimal models from the AGT representation
for non-degenerate fields.

\subsection{Combinatorial representation}

Using the idea of the orthogonal basis for minimal models
we must drop out all  basis elements belonging to invariant submodules.
Even though the explicit construction of the vectors is not known we can use \eqref{4pointBlock} to find for
which  elements of the AGT basis the norm vanishes
\be
\langle \vec{\lambda} |\vec{\lambda}\rangle=0\,.
\ee
This leads us to some  additional restrictions on the form  of
Young diagrams parametrising basis elements in the irreducible representations
of minimal models.

\begin{prop}
\label{FF1}
Consider a degenerate module parameterized by $P_{n,m}$ \eqref{Liouv2}. Function
$Z_v(\Delta_{n,m}, \lambda_1, \lambda_2)$ \eqref{Zv} is not equal to zero provided that Young diagrams
are ordered as
\be
\label{ineqproplength}
(\lambda_1)_i \geq (\lambda_2)_{i+m-1} - n+1\;,
\ee
where $(\lambda_{\alpha})_i$ are lengths of $i$-th rows of Young  diagrams $\lambda_\alpha$, $\alpha = 1,2$.
\end{prop}

The proof is relegated to Appendix \bref{sec:appendixA}. Next, we consider a minimal model $\cM_{p,p^\prime}$ so
that parameter $b$ is given by formula \eqref{CenChaCFT}.
It follows that the function $Z_v$ can have additional zeros that restrict the  basis.
One proves \footnote{The analogous theorem has been established in \cite{FFJMM} within  the representation theory of
the toroidal algebra $\mathbf{gl}_{1}(\mathrm{q}_1,\mathrm{q}_2)$.}
\begin{prop}\label{VirMMregion}
Consider a degenerate module with dimension $\Delta_{n,m}$ in a minimal model $\cM_{p,p^\prime}$.
Function $Z_v(p,p^\prime|\Delta_{n,m}, \lambda_1, \lambda_2)$ is not equal to zero provided the set of Young
diagrams belongs to the region
\be
\label{ineqproplength2}
R_{n,m}^{(p,q)}:\quad(\lambda_{\alpha})_i \geq (\lambda_{\alpha+1})_{i+m_\alpha-1} - n_\alpha+1\;, \qquad \alpha = 1,2\;,
\ee
where $(\lambda_{\alpha})_i$ are lengths of $i$-th rows of Young  diagrams $\lambda_\alpha$,
and $(n_1, m_1) = (n,m)$ and $(n_2, m_2) = (p-n, p^\prime -m)$. We use the identification $\lambda_3 = \lambda_1$.
\end{prop}
The proof is relegated to Appendix \bref{sec:appendixA}. From the Proposition \bref{VirMMregion} we derive  the
following explicit representation for the 4--point conformal block in the
Virasoro minimal models
\begin{equation}
\label{4pointBlockMM}
\mathcal{B}(P_{n_i,m_i};P_{n,m};x)=(1-x)^{-\nu}\sum_{N=0}^{\infty}\!
\,x^{N} \sum^{|\vec{\lambda}|=N}_{\vec{\lambda} \in R_{n,m}^{(p,q)}}\frac{Z_{f}(\mu_i,a;\vec \lambda)}{Z_{v}(a;\vec \lambda)}\,,
\end{equation}
where $P_{n_i,m_i}$ denote external conformal dimensions,
$P_{n,m}$ denote internal one and the summation region $R_{n,m}^{(p,q)}$ is defined in
\eqref{ineqproplength2}. An analogues  result on the conformal blocks in the Virasoro minimal models is  obtained in
Ref. \cite{Bershtein:2014qma}.

\section{Testing AGT for Virasoro minimal models}

In this section we consider a 4--point conformal block
$\mathcal{B}(\Delta_{1,n_j};x)$ with at least
one degenerate field $\Phi_{1,2}$.
This function satisfies the null vector equation
which turns out to be the Riemann equation \cite{BPZ}
\begin{equation}
\label{RiemCFT}
\frac{d^2\mathcal{B}}{d x^2}+\Bigl(\frac{1-\alpha-\alpha'}{x}+
\frac{1-\gamma-\gamma'}{x-1}\Bigr)\frac{d\mathcal{B}}{dx}
+\Bigl(\frac{\alpha\alpha'}{x^2} +\frac{\gamma\gamma'}
{(x-1)^2}+\frac{\beta\beta'-\alpha\alpha'-
\gamma\gamma'}{x(x-1)} \Bigr)\mathcal{B}=0\,,
\end{equation}
with the parameters
\begin{eqnarray}
\label{notat}
&&\alpha=\frac{n_1-1}{2}\kappa\,,\hspace{1.5cm} \alpha'=1-\frac{n_1+1}{2}\kappa\,,\nonumber\\
&&\beta=\frac{2-n_3}{2}\kappa\,,\hspace{1.5cm} \beta'=\frac{n_3+2}{2}\kappa-1\,,\\
&&\gamma=\frac{n_2-1}{2}\kappa\,,\hspace{1.5cm} \gamma'=1-\frac{n_2+1}{2}\kappa\,\nonumber.
\end{eqnarray}
Here, the parameter $\kappa$ is related to the labels
$(p,p')$ of the minimal model as $\kappa=p/p'$.
The equation~\eqref{RiemCFT} has two linearly independent solutions
with  power-law behavior at $x=0$.
They  correspond to two conformal blocks in the S-channel
\begin{equation}
\begin{aligned}
\label{SChanCFT}
&\mathcal{B}_1(x)=x^{\alpha}\,(1-x)^{\gamma}
{}_2F_1(\alpha\,+\beta+\gamma,\alpha+\beta'+\gamma, 1+\alpha-\alpha'; x )\ ,
\\
&\mathcal{B}_2(x)=x^{\alpha'}(1-x)^{\gamma}
{}_2F_1(\alpha'+\beta+\gamma,\alpha'+\beta'+\gamma,1+\alpha'-\alpha; x )\ .
\end{aligned}
\end{equation}
In what follows we reproduce the above hypergeometric functions in the form of the diagrammatic decomposition
\eqref{4pointBlock}, \eqref{4pointBlockMM} applied to the Lee-Yang model.

\subsection{Lee-Yang model}

For the Lee-Yang model ${{\cal M}}_{2,5}$ there is
only one non-trivial primary field in the Kac table $\Phi_{1,2}(=\Phi_{1,3})$
with conformal weight $-1/5$. Due to the fusion rules \eqref{VirFusion}, there are two possible intermediate
channels $\Phi_{1,3}$ and $\Phi_{1,1}$, the corresponding 4--point
conformal blocks are (we omit four external parameters $\Delta_{1,2}$):
\begin{equation}\label{M25confblock}
\begin{aligned}
&\mathcal{B}(\Delta_{1,1};x)=(1-x)^{1/5} {}_2F_1(2/5,3/5,6/5,x)=1 - \frac{x^2}{55} -\frac{x^3}{55} - \frac{9 x^4}{550 }
- \frac{4 x^5}{275}+...\;,\\
&\mathcal{B}(\Delta_{1,3};x)= (1-x)^{1/5} {}_2F_1(1/5,2/5,4/5,x)=1 - \frac{x}{10} - \frac{4 x^2}{75} - \frac{9 x^3}{250}
 - \frac{962 x^4}{35625} +... \;.
\end{aligned}
\end{equation}
In this example,  eq. \eqref{4pointBlockMM} restricts the sum over Young diagrams to be of a general form $(\lambda,\varnothing)$
and $(\varnothing,\lambda)$ for the expansions coefficients $F^{(N)}(\Phi_{1,1})$ and $F^{(N)}(\Phi_{1,3})$ defined in \eqref{4pointBlock}.

At level 1, we have $(\tableau{1},\varnothing)$ and $(\varnothing,\tableau{1})$. The corresponding contributions
are
\begin{eqnarray}
F_{\Delta_{1,1}}^{(1)}=-\frac{(a + \mu_1) (a + \mu_2) (a + \mu_3) (a + \mu_4)}{2 a \epsilon_1 \epsilon_2 (2 a + \epsilon_1 + \epsilon_2)}\,,\nonumber\\
F_{\Delta_{1,3}}^{(1)}=-\frac{(a - \mu_1) (a - \mu_2) (a - \mu_3) (a - \mu_4)}{2 a \epsilon_1 \epsilon_2 (2 a - \epsilon_1 - \epsilon_2)}\,,
\end{eqnarray}
where $\epsilon_1=b$ and $\epsilon_2=b^{-1}$. With \eqref{AGTparam} one can check that ($\nu = -1/5$)
\begin{equation}
\label{checks}
F_{\Delta_{1,1}}^{(1)}+\nu=0\,, \qquad F_{\Delta_{1,3}}^{(1)}+\nu=-\frac{1}{10}\,.
\end{equation}

At level 2, we have $(\tableau{2},\varnothing)$
and $(\varnothing,\tableau{2})$.
The corresponding contributions can be easily derived,
and one can check that
\begin{equation}
F_{\Delta_{1,1}}^{(2)}+\nu F_{\Delta_{1,1}}^{(1)}+\frac{\nu(\nu+1)}{2}=-\frac{1}{55}\,, \qquad
F_{\Delta_{1,3}}^{(2)}+\nu F_{\Delta_{1,3}}^{(1)}+\frac{\nu(\nu+1)}{2}=-\frac{4}{75}\,.
\end{equation}

Let us compare these results with the diagrammatic decomposition at the arbitrary  level $N$. Recall that
intermediate fields are  $\Phi_{1,1}$ and $\Phi_{1,3}$, while all external fields are $\Phi_{1,2}$.

\vspace{-3mm}

\paragraph{Intermediate field $\Phi_{1,1}$.} Following the general consideration  of Section \bref{sec:3.1}
we conclude that diagrams falling out of the Nekrasov decomposition correspond to zeros
of functions $\phi(a|s) + \mu_i =  0$ or $\phi(-a|s) + \mu_i =  0$, cf. \eqref{Zf}.

One obtains that $\Delta_{1,1} =0 $ and  respective $P_0 = \pm Q/2$. We choose $P_0 = Q/2 \equiv a$,
see \eqref{AGTparam}. Also,
\be
\ba{c}
\dps
\mu_1 = \mu_3 = \frac{Q}{2} - 2P_{1,2} = a - 2P_{1,2}\;,
\qquad
\mu_2 = \mu_4 = \frac{Q}{2}  = a\;.
\ea
\ee
Then, consider a factor $\phi(-a|s)+\mu_2 = \phi(-a|s)+a =  0$ in the product over cells of the second Young diagram
$\lambda_2$, cf.  \eqref{Zf}. It follows that this equation reduces to
$b^{-1}(i-1)+b(j-1) = 0$, and the zeros are given by $i=j=1$. It follows that $\lambda_2 = \varnothing$.
On the other hand, consider a factor $\phi(a|s)+\mu_1 = 2a-2P_{1,2} +b^{-1}(i-1)+b(j-1)=  0$ in the product over cells of
the first Young diagram $\lambda_1$, cf.  \eqref{Zf}. The resulting equation is $b^{-1}(i-2)+b(j-1) = 0$ and the zeros are given by $i=2$ and
$j=1$. It follows that $\lambda_1 $ is  an arbitrary length $N$ row, where $N$ is the level.

We conclude here that a decomposition involves pairs of diagrams of the form
\be
\label{dede}
(\lambda_1 = \text{ a row of length}\; N, \;\; \lambda_2 = \varnothing).
\ee
Provided these facts  the diagrammatic decomposition yields   the final formula
\be
F_{\Delta_{1,1}}^{(N)} = \frac{1}{N!}\prod_{n=1}^N \frac{\big(b(n-1)-b^{-1}\big)\big(bn+b^{-1}\big)}{b\big(b(n+1)+2b^{-1}\big)}\;.
\ee
The right-hand-side is given by $N$-th expansion coefficient of the  hypergeometric function \eqref{M25confblock}.
One can check that $F_{\Delta_{1,1}}^{(1)} = -1/5$ that agrees with \eqref{checks}.

\vspace{-3mm}

\paragraph{Intermediate field $\Phi_{1,3}$.} Quite analogously, we consider the case of another intermediate dimension.
One obtains that $\Delta_{1,2}$ corresponds to $P_0 = a= P_{1,3} = \frac{1}{2}(b+3b^{-1})$. An equivalent form
of $a$ reads $a = Q/2 +b^{-1}$ or $Q/2 = a-b^{-1}$. Also,
\be
\ba{c}
\dps
\mu_1 = \mu_3 = a - 2P_{1,2} - b^{-1}\;,
\qquad
\mu_2 = \mu_4 = \frac{Q}{2}  = a-b^{-1}\;.
\ea
\ee
Consider then a factor $\phi(-a|s)+\mu_2 = \phi(-a|s)+a - b^{-1} =  0$ in the product over cells of the second Young
diagram $\lambda_2$, cf.  \eqref{Zf}. It follows that this equation reduces to $b^{-1}(i-2)+b(j-1) = 0$, and the zeros are given by $i=2$, $j=1$.
It follows that $\lambda_2 = $ an arbitrary length row.
On the other hand,
consider a factor $\phi(a|s)+\mu_1 = 2a-2P_{1,2} - b^{-1} + b^{-1}(i-1)+b(j-1)=  0$ in the product over cells of the first Young diagram
$Y_1$, cf.  \eqref{Zf}.
The resulting equation is $b^{-1}(i-1)+b(j-1) = 0$ and the zeros are given by $i=j=1$. It follows that $\lambda_1 = \varnothing$
and therefore $\lambda_1 $ is  an arbitrary length $N$
row, where $N$ is the level.

We conclude here that a decomposition involves pairs of diagrams of the form
\be
(\lambda_1 = \varnothing, \;\; \lambda_2 = \text{ a row of length}\; N).
\ee
Provided these facts  the diagrammatic decomposition yields the final formula
\be
F_{\Delta_{1,3}}^{(N)} = \frac{1}{N!}\prod_{n=1}^N \frac{\big(b(n-1)-b^{-1}\big)\big(b(n-2)-3b^{-1}\big)}{b\big(b(n-1)-2b^{-1}\big)}\;.
\ee
These ratios  are again expansion coefficients of the  hypergeometric function in \eqref{M25confblock}.

\section{Generalization for $\cW_N$ minimal models}
\label{sec:W_N}
%\subsection{$W_N$ Minimal Models basic notations}

In this section we are  interested in unitary $\mathcal{W}_N$ minimal models $\mathcal{M}_{p,p+1}(N)$ with the central charge
\begin{equation}
c = (N-1) \Bigl( 1 - \frac{N(N+1)}{p(p+1)} \Bigr) \,.
\end{equation}
The primary fields are labelled  by two $\widehat{\mathfrak{sl}}_N$
weights   $\Lambda_+=\sum_{s=1}^{N-1} (m_s-1) \omega_s$
and $\Lambda_-=\sum_{s=1}^{N-1} (n_s-1) \omega_s$,
$(m_s, n_s \in \mathbb{Z}_{> 0})$, where $\omega_s$ are the fundamental weights of the Lie algebra $\mathfrak{sl}_{N}$,  and
\be
\label{52}
\sum_{s=1}^{N} m_s = p, \qquad \sum_{s=1}^{N} n_s  = p+1\;,
\ee
where $m_N$ and $n_N$ are defined by the above formulas.
In the Liouville-like parametrization we write
$\Phi_{P}$, where the vector $P=(P^{(1)},...,P^{(N-1)})$, and
\be
\label{53}
P = Q \rho- \textbf{a},\qquad
a_{m,n}=-m b-n b^{-1}, \qquad Q=b+b^{-1} \quad \text{and}\quad b^2=-\frac{p}{p+1}\,,
\ee
where $\rho$ is the Weyl vector (a half-sum of positive roots).
Unlike the Virasoro case, in a $\cW_N$ theory the conformal blocks are not fixed by conformal
and $\cW_N$ invariance \cite{BW}.
The bootstrap program for $k$--point correlation functions
can be performed only if the charges of $k-2$ fields are proportional to the first fundamental weight $\omega_1$ of
the Lie algebra $\mathfrak{sl}_N$ \cite{Fateev:2007ab}.
We consider  the correlation functions of this kind
\be
\langle\Phi_{P} (z_1)\Phi_{a_2}(z_2) \Phi_{a_3}(z_3)... \Phi_{a_{k-1}}(z_{k-1})
\Phi_{\hat P} (z_k)\rangle\;,
\ee
where parameters in points $z_2, ..., z_{k-1}$ correspond to degenerate representations of minimal models described above, while fields in $z_1$ and $z_k$ are general primary fields of the minimal models.
It is convenient to change the variables
\begin{equation}
    z_{i+1}=q_{i}q_{i+1}\cdots q_{k-3}\quad\text{for}\;\;\;i = 1, ..., k-3\;.
\end{equation}
The holomorphic dependence of the correlation functions is encoded in the conformal block
functions
\be
\label{confB}
\mathcal{B}(q_1,...,q_{k-3}|P,P_1,...,P_{k-3},\hat P|a_2,...,a_{k-1})\;,
\ee
where  the momenta $P_1,...,P_{k-3}$ correspond to the fields in the intermediate channels of the conformal block decomposition.
Recall that in the minimal models, the weights of all external and intermediate fields are related by fusion
rules.

In \cite{FFJMM,FFJMM2} there was proposed the orthogonal basis
for modules of the toroidal algebra in some special limits. This basis is labelled by the special sets of  $N$-tuples of Young diagrams.
It was shown that the characters
of these diagrams coincide with the characters of $\cW_N$ minimal  models up to a contribution related to
the presence of extra Heisenberg algebra. It follows  that the found
basis should define AGT basis in the highest weight representations of the $\cH\otimes \cW_N$ algebra \cite{Litvinov}
(see also \cite{Mironov:2013oaa})  restricted for the minimal models thereby giving rise to the AGT representation
for conformal blocks in $\cW_N$ minimal models.
The basis vectors are enumerated by $N$-tuples of Young diagrams
with some additional restrictions formulated below.

We conjecture
the following explicit form of the $k$--point conformal block in $\cW_N$
minimal models $\cM_{p,p+1}$ ($p\geq N-1$).
In this case  the conformal block \eqref{confB} for non-degenerate parameters is related to the instanton part of the Nekrasov partition function for the quiver gauge theory
and can be written explicitly \cite{Litvinov,Wyl}
\begin{equation}\label{conformal-block-refined}
 \mathcal{B} {=}
      \prod_{j=1}^{k-3}\prod_{l=j}^{k-3}(1-q_{j}\cdots  q_{l})^{-a_{j+1}(Q-a_{l+2}/N)}\,\,
    \mathcal{F}\;,
\end{equation}
and
\begin{equation}\label{conformal-block-refined-explicit}
    \mathcal{F}=1+\sum_{\vec{j}}q_{1}^{j_{1}}q_{2}^{j_{2}}\dots q_{k-3}^{j_{k-3}}\,\mathbb{Z}_{\vec{j}}\;,
\end{equation}
where $\vec{j} = (j_1, ..., j_{k-3})$,  with the coefficients
\begin{multline}
\label{conformal-block1}
   \mathbb{Z}_{\vec{j}}=
   \sum^{|\vec{\lambda}|=r_i}_{\vec{\lambda}_{i}\in \mathcal{R}_{\textbf{m}_i,\textbf{n}_i}}
   \mathbb{N}^{-1}(\vec{\lambda}_{1},P_{1})\dots \mathbb{N}^{-1}(\vec{\lambda}_{k-3},P_{k-3})\times\\\times
   \mathbb{F}^{\vec{\lambda}_{1}}_{\scriptscriptstyle{\varnothing}}(a_{2},P_{1},P)
   \mathbb{F}^{\vec{\lambda}_{2}}_{\vec{\lambda}_{1}}(a_{3},P_{2},P_{1})
   \dots
   \mathbb{F}^{\vec{\lambda}_{k-3}}_{\vec{\lambda}_{k-4}}(a_{k-2},P_{k-3},P_{k-4})
   \mathbb{F}^{\scriptscriptstyle{\varnothing}}_{\vec{\lambda}_{k-3}}(a_{k-1},\hat{P},P_{k-3})\;.
\end{multline}
Here  $\vec{\lambda_i}=(\lambda^{(1)}_i,...,\lambda^{(N)}_i)$  are $N$-tuples of Young diagrams,
and  index $i=1,...,k-3$ enumerates intermediate channels.
Each component of the vectors $\lambda^{(s)}_i$ is a finite integer partition. We define the norms  as
a total number of boxes  in the Young diagram representation
$|\vec{\lambda}_i|=\sum_{s=1}^N  |\lambda^{(s)}_i|$.
The norms are  $\mathbb{N}(\vec{\lambda}, P)=\mathbb{F}^{\vec{\lambda}}_{\vec{\lambda}}(0,- P, P)$, and
the general matrix element is given by \cite{Litvinov}
\begin{eqnarray}
\label{Zbif-def}
\mathbb{F}_{\vec{\lambda}^{\prime}}^{\vec{\lambda}}(a,P,P^{\prime})=\!\prod\limits_{i,j=1}^{N}\!\prod\limits_{t^\prime\in\lambda_{i}^{\prime}}
(Q-E_{\lambda_{i}^{\prime},\lambda_{j}}(x_{j}-x_{i}^{\prime}|t^\prime)-a/N)\!\prod\limits_{t\in\lambda_{j}}
(E_{\lambda_{j},\lambda_{i}^{\prime}}(x_{i}^{\prime}-x_{j}|t)-a/N),
\end{eqnarray}
where $x_j = (h_j , P)$ (vectors $h_i$  are the weights of the first fundamental representation
of $\mathfrak{sl}_{N}$ with the the highest weight $\omega_1$, \textit{i.e.}
$h_i = \omega_1 - e_1 - ... - e_{i-1}$, where $e_k$ are simple roots, and
$h_i h_j = 1- \frac{1}{N}\delta_{ij}$), while
 the function $E_{\lambda,\mu}\!(x\bigl|t\bigr)$ is defined in \eqref{E-def}.
 \footnote{Strictly speaking, formula \eqref{Zbif-def} is valid for the case $N\geq 3$ only.
For  $N=2$ one should use a different normalization.}

Extending the Proposition \bref{VirMMregion} we conjecture
that the summation in formula \eqref{conformal-block1} is further restricted by the following region
\begin{equation}
\label{summation}
\mathcal{R}_{\textbf{m},\textbf{n}}=
\left\{\vec \lambda \,\,\big|\,\,  \big(\lambda^{(s)}\big)_j\geq
\big(\lambda^{(s+1)}\big)_{j+m_s-1}\!-n_s+1, \text{ where } s=1,...,N\!\,,\;\; j\in \mathbb{Z}_{>0}  \right\}\,,
\end{equation}
where $\lambda^{(N+1)}\equiv \lambda^{(1)}$, while $m_s$ and $n_s$ are components of extended $\mathfrak{sl}_{N}$ weight vectors
$\textbf{m}  = (m_1, ..., m_{N})$ and $\textbf{n} = (n_1, ..., n_{N})$, see \eqref{52}.

\section{Conclusions}

In this paper we studied application of the AGT correspondence to
minimal models of Virasoro and $\cW_N$ algebras. We used the
conjecture (supported by the comparison of the characters of the
corresponding representations) that the chiral $\cW_N$ algebra
appears in the conformal limit from the toroidal algebra
$\mathbf{gl}_{1}(\mathrm{q}_1,\mathrm{q}_2)$ \cite{FFJMM,FFJMM2}. This connection
reveals a nice integrable structure of degenerate modules of the
algebra $\mathcal{A}_N=\mathcal{H}\otimes \cW_N$. It appears for the $\cW_N$
central charge corresponding to the minimal models $\cM_{p,p^\prime}$
arising  from $\mathbf{gl}_{1}(\mathrm{q}_1,\mathrm{q}_2)$ once the parameters
are constrained by the following wheel condition $\mathrm{q}_1^p \mathrm{q}_2^{p^\prime}=1$.
In particular, $\mathcal{A}_N$ module inherits some (reduced)
orthogonal basis defined naturally in $\mathbf{gl}_{1}(\mathrm{q}_1,\mathrm{q}_2)$
modules. Hence, it is natural to use this basis for the evaluation
of the $\cW_N$ conformal block functions. We have checked that
using this basis for Virasoro minimal models we get the AGT-like
representation for the conformal blocks of minimal models once the
consequences of emergence of invariant subspaces in the degenerate
representations is taken into account.

Our  main result is the explicit expression for the conformal
blocks in the Virasoro minimal models \eqref{4pointBlockMM} and
the conjecture on  the form  of  AGT representation for
$\cW_N$ conformal blocks \eqref{conformal-block1} -
\eqref{summation}. The difference from the original  AGT expression
for nondegenerate representations is encoded in additional
restrictions on the summation region over Young diagrams.

\vspace{5mm}

\noindent \textbf{Acknowledgements.} We are grateful to A.
Belavin, B. Feigin, and A. Litvinov for  useful discussions.  V.B.
would like to thank G. Mussardo, SISSA and K. Narain, ICTP,
Trieste, Italy for hospitality during his visits in 2013.
The work
of K.A. is partially supported by the RFBR grant No 14-02-01171.
The work of V.B.
is partially supported by the RFBR grant No 12-01-00836-a.

%\vspace{5mm}

%%%%%%%%%%%%%%%%%%%%%%%%%%%%%%%%%%%%%%%%%%%%%%%%%%%%%%%%%%%%%%%%%%%%%%%%
%%%%%%%%%%%%%%%%%%%%%%%%%%%%%%%%%%%%%%%%%%%%%%%%%%%%%%%%%%%%%%%%%%%%%%%%
%%%%%%%%%%%%%%%%%%%%%%%%%%%%%%%%%%%%%%%%%%%%%%%%%%%%%%%%%%%%%%%%%%%%%%%%

\appendix

\section{Proof of the propositions}
\label{sec:appendixA}

In order to analyse zeros of functions $Z_f$ and $Z_v$ we use simple  method described below. It turns out
that for particular values of external/internal dimensions and/or the central charge
these functions take the form of a vector on the $(b^{-1}, b)$ plane with integer-valued
coordinates
\be
\label{master_zero}
F(s) \equiv  (j - K)b\, \pm (i-L) b^{-1} = 0\;,
\ee
where $K,L$ are some integers and $(i,j)$ are coordinates of a cell $s\in \lambda_1$ or $s\in \lambda_2$.
For general value of the Liouville coupling $b$ there is a unique solution $ i = L$ and $j = K$
which defines a cell where function $F(s) = 0$. Obviously, only positive numbers $K,L$ make sense
so one can define a set of "admissible" diagrams $(\lambda_1, \lambda_2)$  which do not contain cells with
coordinates $(L,K)$. In what follows ($\stackrel{-}{x}, \stackrel{-}{y}$)  and ($\stackrel{=}{x}, \stackrel{=}{y}$)
denote coordinates of excluded cells (those that give zeros of the functions) in Young diagrams $\lambda_1$ and $\lambda_2$, respectively.

For the minimal models $\cM_{p,p^\prime}$ the Liouville coupling takes particular value \eqref{CenChaCFT}
and therefore  equation \eqref{master_zero} allows for more (infinitely many) solutions, namely,
$i = L \pm \alpha p^\prime$ and $j = K + \alpha p$. Here arbitrary parameter $\alpha \in \mathbb{Z}$
because $p$ and $p^\prime$ are coprimes. Note that the case $\alpha = 0$ reproduces zeros described
above for general Liouville coupling $b$. It follows that for the minimal models more zeros appear but sometimes
new zeros are "weaker" than those for  $\forall \,b$. It is worth noting that actually there are
infinitely many new zeros but generally values their coordinates are restricted from below
by minimal values that define the form of admissible diagrams.

\vspace{-3mm}
\paragraph{Proof of the Proposition \bref{FF1}.}

First of all we note that the terms with $E_{\lambda_\alpha, \lambda_{\alpha}}\!\bigl(x\bigl|s\bigr)$ are not
equal to zero. The only source of zeros is provided by terms containing $E_{\lambda_{\alpha},\lambda_{\beta}}\!\bigl(x\bigl|s\bigr)$ with the pair of different Young diagrams.

We start  with empty diagram $\lambda_2$ and subsequent increase a number of
its rows. Another trick is to consider cases $\Delta_{1,m}$ and $\Delta_{n,1}$ separately so that
the general case of $\Delta_{n,m}$ is a straightforward combination of the previous ones.
On the plane $(b^{-1}, b)$ we have
\be
\label{EY1Y2}
E_{\lambda_1,\lambda_2}\!\bigl(2a\bigl|s\bigr) = 0\; : \qquad
b^{-1}(\stackrel{-}{h}_j - \stackrel{-}{i}+m+1) - b(\stackrel{=}{k}_i - \stackrel{-}{j}-n) = 0
\ee
or
\be
\label{EY1Y2cell}
\stackrel{-}{i}\; =\;  \stackrel{-}{h}_j + m+1\;,
\qquad
\stackrel{-}{j} \; =\; \stackrel{=}{k}_i - n\;.
\ee
and
\be
\label{EY2Y1}
E_{\lambda_2,\lambda_1}\!\bigl(-2a\bigl|s\bigr) = 0 \; : \qquad
b^{-1}(\stackrel{=}{h}_j - \stackrel{=}{i}-m+1) - b(\stackrel{-}{k}_i - \stackrel{=}{j}+n) = 0
\ee
or
\be
\label{EY2Y1cell}
\stackrel{=}{i}\; =\;  \stackrel{=}{h}_j - m+1\;,
\qquad
\stackrel{=}{j} \; =\; \stackrel{-}{k}_i +n\;.
\ee

Consider diagrams $(\lambda_1, \varnothing)$ with any $\lambda_1$, and show that these do not produce zeros of $Z_v$.
Indeed, in this case $\stackrel{=}{k}_i = \stackrel{=}{h}_j = 0 $ so that from \eqref{EY1Y2cell}
one derives $\stackrel{-}{j} = -n<0$. Equation \eqref{EY2Y1} is absent in this case. Therefore,
we conclude that  pairs $(\lambda_1, \varnothing)$ are admissible.

%\vspace{5mm}

\noindent \textit{Dimension $\Delta_{1,m}$.} Consider pairs of  diagrams $(\lambda_1, \lambda_2)$, where
$\lambda_2  = (N_1, 0, 0,...)$ is a row of arbitrary length $N_1$.
Coordinates of a cell in a row are $(1, \stackrel{=}{j})$, where $\stackrel{=}{j} = 1,...,N_1$.
Our aim is to show that absence of zeros imposes constraints on the form of  diagram $\lambda_1$.

From
equation \eqref{EY2Y1cell} one obtains coordinates of a cell that produces a zero,
\be
\label{tomsk1}
\stackrel{=}{i}\; =\;  \stackrel{=}{h}_j - m+1\;,
\qquad
\stackrel{=}{j} \; =\; \stackrel{-}{k}_i +1\;.
\ee
In the case of $\lambda_2=$  row one derives from the first equation above that $\stackrel{=}{h}_j  = m$,
and it follows that $m =1$ to have a solution. Another way around, it implies that zeros appear for those
$\lambda_2$  that have at least $m$ rows, \textit{i.e.}, $\exists \stackrel{=}{j}$ such that $\stackrel{=}{h}_j\; \geq \; m$.

For $m=1$ the second equation above says that in order to have a zero
a first row of $\lambda_1$ is to be of length $\stackrel{-}{k}_1 \;\leq\; N_1-1$. One concludes that for $m\neq 1$
zeros are absent for any $\lambda_1$, while for $m=1$
zeros are absent for pairs of diagrams with lengths subject to
\be
\label{tomsk2}
\stackrel{-}{k}_1\; \geq \;  \stackrel{=}{k}_1\;.
\ee

Consider then pairs of  diagrams $(\lambda_1, \lambda_2)$, where $\lambda_2$ is an arbitrary diagram with $m$ rows
of ordered lengths $N_1 \geq  N_2 \geq  ... \geq N_m$. Substituting $\stackrel{=}{i}=1$ and
$\stackrel{=}{j} = 1,..., N_m$ to the first equation in \eqref{tomsk1}
gives solution, $1 = m - m +1$. The second equation in \eqref{tomsk1} takes the form $\stackrel{=}{j} =
\stackrel{=}{k}_1 +1$ which defines $\stackrel{=}{k}_1 = 0,1,..., N_m-1$, while lengths $\stackrel{=}{k}_\alpha$,
where $\alpha \geq 2$ are arbitrary but no bigger than $\stackrel{=}{k}_1$. One concludes that zeros
are absent for pairs of diagrams with lengths subject to the following inequality
\be
\label{tomsk3}
\stackrel{-}{k}_1\; \geq \;  \stackrel{=}{k}_m\;,
\ee
which is obviously generalizes \eqref{tomsk2} to arbitrary value of $m$.

As the next step one considers an arbitrary diagram $\lambda_2$ with $m+l$ rows, where $l=1,2,...$. The diagram
$\lambda_2$ naturally splits in two subdiagrams $\lambda_2 = \lambda^\prime_2\oplus \lambda_2^{\prime\prime}$, where the first factor
is a diagram composed of first $m-1$ rows of $\lambda_2$, while the second one is a diagram composed of the remaining rows of
$\lambda_2$. Considering equations \eqref{tomsk1} one shows that zeros are absent when
$\lambda_2^{\prime\prime} \subseteq \lambda_1$.
Equivalently,
\be
\label{tomsk4}
\stackrel{-}{k}_i\; \geq \;  \stackrel{=}{k}_{m+i-1}\;.
\ee
This inequality completely describes admissible pairs of diagrams $(\lambda_1, \lambda_2)$ in the case of
dimension $\Delta_{1,m}$.

%\vspace{5mm}

\noindent\textit{Dimension $\Delta_{n,1}$.} Consider pairs of  diagrams $(\lambda_1, \lambda_2)$,
where $\lambda_2  = (N_1, 0, 0,...)$ is a row of arbitrary length $N_1$.
From
equation \eqref{EY2Y1cell} one obtains coordinates of a cell that produces a zero,
\be
\label{tomsk5}
\stackrel{=}{i}\; =\;  \stackrel{=}{h}_j\;,
\qquad
\stackrel{=}{j} \; =\; \stackrel{-}{k}_i +n\;.
\ee
In the case of $\lambda_2=$ the first equation is automatically satisfied for $\stackrel{=}{j} = 1,..., N_1$.
The second equation says that zeros appear when the first row of $\lambda_1$ is of length less than $N_1-n$.
This is to say that zeros are absent for pairs of diagrams with lengths subject to
\be
\label{tomsk6}
\stackrel{-}{k}_1\; \geq \;  \stackrel{=}{k}_1 - n+1\;.
\ee
This inequality naturally generalizes to the case of $\lambda_2$ with any number of rows. Namely,
\be
\label{tomsk7}
\stackrel{-}{k}_i\; \geq \;  \stackrel{=}{k}_i - n+1\;.
\ee

\vspace{5mm}

\noindent \textit{Dimension $\Delta_{n,m}$.} To find admissible diagrams in  the case of arbitrary dimension $\Delta_{n,m}$
one simply combines previously considered cases of $\Delta_{1,m}$ and $\Delta_{n,1}$ to obtain
formula \eqref{ineqproplength}
\be
\label{tomsk8}
\stackrel{-}{k}_i\; \geq \;  \stackrel{=}{k}_{i+m-1} - n+1\;.
\ee
In particular, this relation implies that in order to produce a zero the second diagram $Y_2$ should
include a rectangle of length $n$ and height $m$.

To conclude the proof one notes that zeros are also contained in $E_{\lambda_2,\lambda_1}\!\bigl(-2a\bigl|s\bigr)-Q = 0$.
Equations that define coordinates of a zero are those in \eqref{EY2Y1cell} but with $m \rightarrow m+1$
and $n \rightarrow n+1$. Admissible diagrams are defined by inequality
$\stackrel{-}{k}_i\; \geq \;  \stackrel{=}{k}_{i+m} - n$ which is weaker than \eqref{tomsk8} though.
Indeed, using a definition of a Young diagram one observes that $\stackrel{=}{k}_{i+m-1} \; \geq \; \stackrel{=}{k}_{i+m}$
which takes \eqref{tomsk8} to the form $\stackrel{-}{k}_i\; \geq \;  \stackrel{=}{k}_{i+m} - n+1
\;>\; \stackrel{=}{k}_{i+m} - n$.

\vspace{-3mm}

\paragraph{Proof of the Proposition \bref{VirMMregion}.}

The proof is similar to that one of Proposition \bref{FF1}. The difference is that more zeros
appear. Indeed, reconsider condition $E_{\lambda_2,\lambda_1}\!\bigl(-2a\bigl|s\bigr) = 0$ from  \eqref{EY2Y1}.
The coordinates of zeros are given by
\be
\label{EY2Y1cell2}
\stackrel{=}{i}\; =\;  \stackrel{=}{h}_j - m+1 - \alpha p^\prime\;,
\qquad
\stackrel{=}{j} \; =\; \stackrel{-}{k}_i +n + \alpha p\;,
\ee
for any  $\alpha \in \mathbb{Z}$ because coordinates are integers, while
$p$ and $p^\prime$ are coprimes. These equations coincide with those in \eqref{EY2Y1cell}
but $n \rightarrow n + \alpha p $ and $m \rightarrow m + \alpha p^\prime $. For $\alpha \in \mathbb{Z}_+ $
one obtains that resulting restrictions of diagrams are weaker than \eqref{ineqproplength}.
To consider the case of  $\alpha \in \mathbb{Z}_- $ one recalls that by definition of minimal models
$n < p$ and $m < p^\prime$. Then one notices that $n + \alpha p <0$ and $m + \alpha p^\prime  <0$
which is to say that zeros are absent. We conclude that condition \eqref{EY2Y1}
does not produce new zeros.

New zeros appear due to condition \eqref{EY1Y2}. In this case coordinates of zeros
are
\be
\label{EY1Y2cell2}
\stackrel{-}{i}\; =\;  \stackrel{-}{h}_j + m+1 - \alpha p^\prime\;,
\qquad
\stackrel{-}{j} \; =\; \stackrel{=}{k}_i - n + \alpha p\;,
\ee
where $\alpha \in \mathbb{Z}$. Zeros are possible for $\alpha \in \mathbb{N}$ only.
The resulting equations coincide with those in \eqref{EY2Y1cell2} provided
$\lambda_1 \leftrightarrow \lambda_2$ and $ n \leftrightarrow (\alpha+1) p -n$ and $ m \leftrightarrow (\alpha+1) p^\prime -m$,
where now $\alpha \in \mathbb{Z}_+$. Repeating arguments below formula \eqref{EY2Y1cell2}
one concludes that equations \eqref{EY1Y2cell2} impose the following restrictions
\be
\label{tomsk9}
\stackrel{=}{k}_i\; \geq \;  \stackrel{-}{k}_{i+(p^\prime -m)-1} - (p-n) +1\;.
\ee
Introducing parameters $(n_1, m_1) = (n,m)$ and $(n_2, m_2) = (p-n, p^\prime -m)$ one
obtains formula \eqref{ineqproplength2} of the proposition.

\providecommand{\href}[2]{#2}\begingroup\raggedright
\addtolength{\baselineskip}{-5pt}
\addtolength{\parskip}{-2pt}
\providecommand{\href}[2]{#2}\begingroup\raggedright


\begin{thebibliography}{10}

\bibitem{BPZ}
A.~A.~Belavin, A.~M.~Polyakov and A.~B.~Zamolodchikov,
``Infinite Conformal Symmetry in Two-Dimensional Quantum Field Theory,''  Nucl.\ Phys.\ B {\bf 241} (1984) 333.


\bibitem{AGT}
  L.~F.~Alday, D.~Gaiotto and Y.~Tachikawa,
  ``Liouville Correlation Functions from Four-dimensional Gauge Theories,''
  Lett.\ Math.\ Phys.\  {\bf 91} (2010) 167  [arXiv:0906.3219 [hep-th]].

\bibitem{Nekrasov}
  N.~A.~Nekrasov,
  ``Seiberg-Witten prepotential from instanton counting,''  Adv.\ Theor.\ Math.\ Phys.\  {\bf 7} (2004) 831  [hep-th/0206161].


\bibitem{Wyl}
  N.~Wyllard,
  ``A(N-1) conformal Toda field theory correlation functions from conformal N = 2 SU(N) quiver gauge theories,''  JHEP {\bf 0911} (2009) 002  [arXiv:0907.2189 [hep-th]].

\bibitem{Mironov:2009qt}
  A.~Mironov and A.~Morozov,
  %``The Power of Nekrasov Functions,''
  Phys.\ Lett.\ B {\bf 680} (2009) 188
  [arXiv:0908.2190 [hep-th]].


\bibitem{Mironov:2009by}
  A.~Mironov and A.~Morozov,
  ``On AGT relation in the case of U(3),''
  Nucl.\ Phys.\ B {\bf 825} (2010) 1
  [arXiv:0908.2569 [hep-th]].


\bibitem{Litvinov}
  V.~A.~Fateev and A.~V.~Litvinov,
  ``Integrable structure, W-symmetry and AGT relation,''  JHEP {\bf 1201} (2012) 051
  [arXiv:1109.4042 [hep-th]].


\bibitem{Belavin:2011pp}
  V.~Belavin and B.~Feigin,
  ``Super Liouville conformal blocks from N=2 SU(2) quiver gauge theories,''  JHEP {\bf 1107} (2011) 079  [arXiv:1105.5800 [hep-th]].


\bibitem{AlbaLitvinov}
  V.~A.~Alba, V.~A.~Fateev, A.~V.~Litvinov and G.~M.~Tarnopolskiy,
  ``On combinatorial expansion of the conformal blocks arising from AGT conjecture,''  Lett.\ Math.\ Phys.\  {\bf 98} (2011) 33  [arXiv:1012.1312 [hep-th]].


\bibitem{Santachiara:2010bt}
  R.~Santachiara and A.~Tanzini,
  %``Moore-Read Fractional Quantum Hall wavefunctions and SU(2) quiver gauge theories,''
  Phys.\ Rev.\ D {\bf 82} (2010) 126006
  [arXiv:1002.5017 [hep-th]].

\bibitem{Estienne:2011qk}
  B.~Estienne, V.~Pasquier, R.~Santachiara and D.~Serban,
  ``Conformal blocks in Virasoro and W theories: Duality and the Calogero-Sutherland model,''
  Nucl.\ Phys.\ B {\bf 860} (2012) 377
  [arXiv:1110.1101 [hep-th]].

\bibitem{Papadodimas:2011pf}
K.~Papadodimas and S.~Raju,
``Correlation functions in holographic minimal models,''
arXiv:1108.3077 [hep-th].

\bibitem{Chang:2011vk}
  C.-M.~Chang and X.~Yin,
  ``Correlators in $\cW_N$ Minimal Model Revisited,''
  arXiv:1112.5459 [hep-th].


\bibitem{FFJMM}

B. Feigin, E. Feigin, M. Jimbo, T. Miwa, E. Mukhin, "Quantum continuous $gl_\infty$: Tensor products of Fock modules and $\cW_N$ characters",
    arXiv:1002.3113.

\bibitem{FFJMM2}

B. Feigin, E. Feigin, M. Jimbo, T. Miwa, E. Mukhin, "Quantum continuous $gl_\infty$: Semi-infinite construction of representations",
    arXiv:1002.3100.


%\bibitem{4dconfblock}
%  Z.~U.~Khandker, D.~Li, D.~Poland and D.~Simmons-Duffin,
%  ``$\mathcal{N}=1$ Superconformal Blocks for General Scalar Operators,''  arXiv:1404.5300 [hep-th].
%

%\bibitem{Bouwknegt:1992wg}
%P.~Bouwknegt and K.~Schoutens,
%{\it W symmetry in conformal field theory},
%Phys.\ Rept.\  {\bf 223} (1993) 183
%{\tt [arXiv:hep-th/9210010]}.
%%%CITATION = PRPLC,223,183;%%
%


\bibitem{AlZ}
A.~Zamolodchikov,
``Higher equations of motion in Liouville field theory,''  Int.\ J.\ Mod.\ Phys.\ A {\bf 19S2} (2004) 510  [hep-th/0312279].


\bibitem{Bershtein:2014qma}
  M.~Bershtein and O.~Foda,
  ``AGT, Burge pairs and minimal models,''  arXiv:1404.7075 [hep-th].



\bibitem{BW}
  P.~Bowcock and G.~M.~T.~Watts,
  ``Null vectors of the W(3) algebra,''  Phys.\ Lett.\ B {\bf 297} (1992) 282  [hep-th/9209105].


\bibitem{Fateev:2007ab}
  V.~A.~Fateev and A.~V.~Litvinov,
  ``Correlation functions in conformal Toda field theory. I.,''  JHEP {\bf 0711} (2007) 002  [arXiv:0709.3806 [hep-th]].


\bibitem{Mironov:2013oaa}
  S.~Mironov, And.~Morozov and Y.~Zenkevich,
  ``Generalized Jack polynomials and the AGT relations for the $SU(3)$ group,''
  JETP Lett.\  {\bf 99} (2014) 109
  [arXiv:1312.5732 [hep-th]].





\end{thebibliography}
\end{document}